\documentclass{article}
\usepackage{graphicx}

\begin{document}

\noindent\underline{Astronomy Reports, 2011, Vol. 55, No. 9, pp. 810--815. DOI: 10.1134/S1063772911090058}

\vskip 5em
\centerline{\LARGE\bf The influence of Galactic aberration on precession \strut} 
\centerline{\LARGE\bf parameters determined from VLBI observations \strut} 
\bigskip
\centerline{\Large\bf Z. M. Malkin}
\bigskip
\centerline{\large Pulkovo Observatory, St. Petersburg, Russia}
\centerline{\large e-mail: malkin@gao.spb.ru}

\vskip 3em
\begin{quotation}
{\bf Abstract.} The influence of proper motions of sources due to Galactic aberration on precession models 
based on VLBI data is determined. Comparisons of the linear trends in the coordinates of the celestial pole 
obtained with and without taking into account Galactic aberration indicate that this effect can reach 20~$\mu$as 
per century, which is important for modern precession models. It is also shown that correcting for Galactic 
aberration influences the derived parameters of low-frequency nutation terms. It is therefore necessary to 
correct for Galactic aberration in the reduction of modern astrometric observations. 
\end{quotation}

\section{Introduction}

With growth in the accuracy of astronomical position
measurements, the requirements for the models
used in the associated astrometric reduction are 
likewise increased. It becomes necessary to take 
into account finer effects influencing the measured 
positions and motions of celestial objects. One of 
these is secular aberration, due to the motion of the 
observer together with the motion of the solar-system 
barycenter (SSB). The existence of secular aberration 
has long been understood. Already 200 years ago, the 
Greenwich astronomer J.~Pond indicated the presence
of a shift in the apparent positions of celestial 
bodies due to the motion of the solar system [1]. 
Secular aberration is a fairly complex phenomenon. 
The full effect is composed of several components, 
corresponding to various components of the velocity 
of the SSB: the motion of the SSB relative to the 
Local Standard of Rest (LSR), the motion of the LSR 
relative to the center of the Galaxy, the motion of 
the Galaxy relative to the Local Group, the motion 
of the Local Group relative to the Local Supercluster,
 the motion of the Local Supercluster etc. The 
motion of galaxies on scales larger than tens of Mpc 
($z>\sim 0.02$) remains poorly studied. Moreover, there 
are fundamental difficulties with such computations, 
associated with the selection of a suitable coordinate 
system [2]. 

Fortunately, the motion of the SSB is linear to 
a high degree, which makes the aberrational shift 
for a given object nearly constant over the comparatively
 short historical period over which accurate 
astronomical observations have been available. Since 
this shift cannot be determined from observations 
or calculated theoretically with sufficient accuracy, it 
has not been usual historically to correct for secular 
aberration in astrometric data reduction, although it 
can comprise up to several arcminutes. At the same 
time, deviations from linear motion of the SSB give 
rise to apparent proper motions of about 5~$\mu$as/yr. 
As was shown in [3, 4], the main contribution to the 
curvature of the motion of the SSB is made by the 
circular rotation of the Galaxy, while the influence of 
the remaining components of the SSB motion is at 
least an order of magnitude smaller. Therefore, in 
view of the smallness of the effect, we will further 
be concerned only with the circular motion of the 
LSR, and will call the corresponding component of 
the secular aberration Galactic aberration (GA). 

Until recently, the proper (apparent) motions of 
celestial bodies due to GA were much smaller than 
the observational accuracy, and were therefore ignored.
However, it became clear as early as the 
1980s that, with time, they would have to be taken 
into account in the reduction of high-accuracy Very 
Long Baseline Interferometry (VLBI) data obtained 
with ground networks [5--7], as well as data from 
space astrometric missions [3, 4, 8, 9]. If not taken 
into account, these aberration proper motions can 
distort observational results with microarcsecond accuracies.
This is relevant first and foremost to the 
determination of secular variations in astronomical 
parameters. 

One example of important such quantities derived 
from long-term series of VLBI observations are parameters
of precession models [10, 11]. The required 
accuracy for these parameters is about one $\mu$as
per century [10]. Currently, these parameters 
are refined using series of coordinates of the celestial 
pole (CP) derived from VLBI data, which are available
starting from 1979 (although it is common to 
omit from consideration the first few years of observations,
which have relatively low accuracy, as we 
have done in the current study). Failure to correct 
the observed motions of radio sources for GA directly
influences the results of observations designed 
to refine precession models. Since this effect depends 
on several factors, such as changes in the observing 
program, the distribution of observed sources on the 
celestial sphere, etc, it is difficult to estimate its magnitude
 theoretically. The current study investigates 
the strength of the influence of GA in practice. 

\section{Influence of GA on radio source proper motions}

The proper motion vector of a celestial object 
due to GA is directed along the Galactocentric acceleration
vector; i.e., toward the center of the 
Galaxy. Its magnitude is given by [3] 
\begin{equation}
A = \frac{V_0 \Omega_0}{c} \,,
\label{eq:const1}
\end{equation}
where $V_0$ is the linear speed of the LSR due to the 
rotation of the Galaxy, $\Omega_0$ is the angular speed of the 
LSR about the Galactic center, and $c$ is the speed 
of light. We will call this quantity the Galactic aberration constant. 

Let us rewrite (1) by expressing $A$ in terms of 
fundamental quantities, determined using Galactic-astronomy methods: 
\begin{equation}
A = \frac{R_0 \Omega_0^2}{c} \,,
\label{eq:const2}
\end{equation}
where $R_0$ is the distance from the SSB to the 
Galactic center. To calculate the GA constant, 
we adopt the mean values $R_0 = 8.2$ kpc and $\Omega_0 = 29.5$ km s$^{-1}$ kpc$^{-1}$ (6.22 mas/yr) 
[13--15], which yields the GA constant $A$ = 5.02~$\mu$as/yr.
The linear rotational speed of the LSR 
about the Galactic center is $V_0$ = 242 km/s$^{-1}$, and the 
rotational period is 208 million years. 

The influence of GA on the coordinates of celestial
bodies in Galactic coordinates can be expressed as [3]: 
\begin{equation}
\begin{array}{rcl}
\mu_l \cos b &=& -A \sin l \,, \\
\mu_b &=& -A \cos l \sin b \,, \\
\end{array}
\label{eq:galactic}
\end{equation}
where $l$ and $b$ are the Galactic longitude and latitude of the object, respectively. 

Since most astrometric calculations, including 
VLBI data processing, are carried out in equatorial 
coordinates, we present the following formulas from 
[12], carrying over the multiplication of $A$ by $1/c$: 
\begin{equation}
\begin{array}{rcl}
\mu_\alpha \cos \delta &=& -A_1 \sin\alpha + A_2 \cos\alpha \,, \\
\mu_\delta &=& -A_1 \cos\alpha \sin\delta - A_2 \sin\alpha \sin\delta \\
           && + A_3 \cos\delta \,, \\
\end{array}
\label{eq:equatorial}
\end{equation}
where 
\begin{equation}
\begin{array}{rcl}
A_1 &=& A \cos\alpha_0 \cos\delta_0, \\ A_2 &=& A \sin\alpha_0 \cos\delta_0, \\ A_3 &=& A \sin\delta_0 \,,
\end{array}
\end{equation}
and $\alpha_0, \ \delta_0$~are the equatorial coordinates of the Galactic center.
With $\alpha_0 = 266.\!^\circ 405100, \ \delta_0 = -28.\!^\circ 936175$,
we obtain $A_1 = -0.28, \ A_2 = -4.39, \ A_3 = -2.43$ $\mu$as/yr.
The proper motions of celestial objects due to GA in both coordinate
systems are shown in Fig.~1. 
Since GA has generally not been taken into account,
it is present in all catalogs of coordinates of 
celestial objects, which are thus apparent coordinates.
To derive true coordinates (corrected 
for GA), the corrections (4) must be subtracted from 
the catalog positions. 

\begin{figure}
\centering
\includegraphics[width=0.48\textwidth,clip]{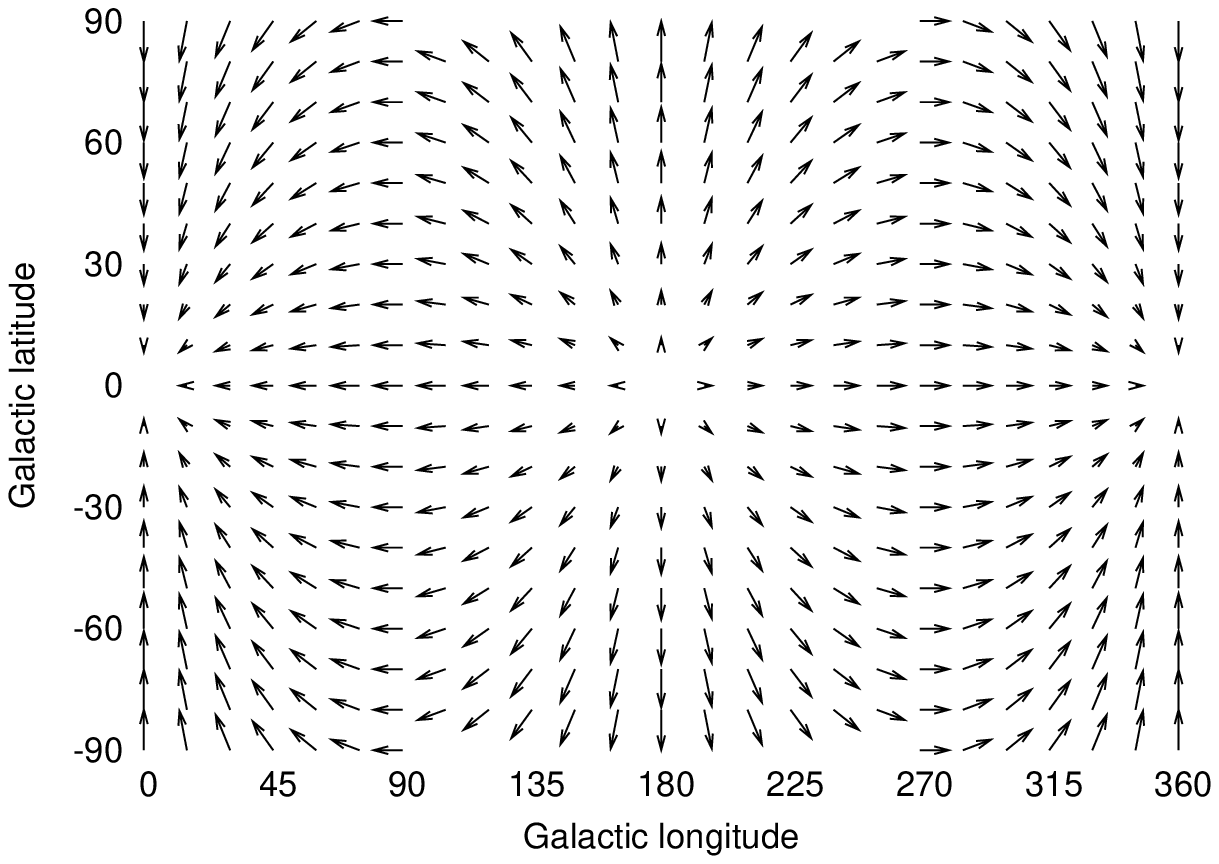}
\hspace{1ex}
\includegraphics[width=0.48\textwidth,clip]{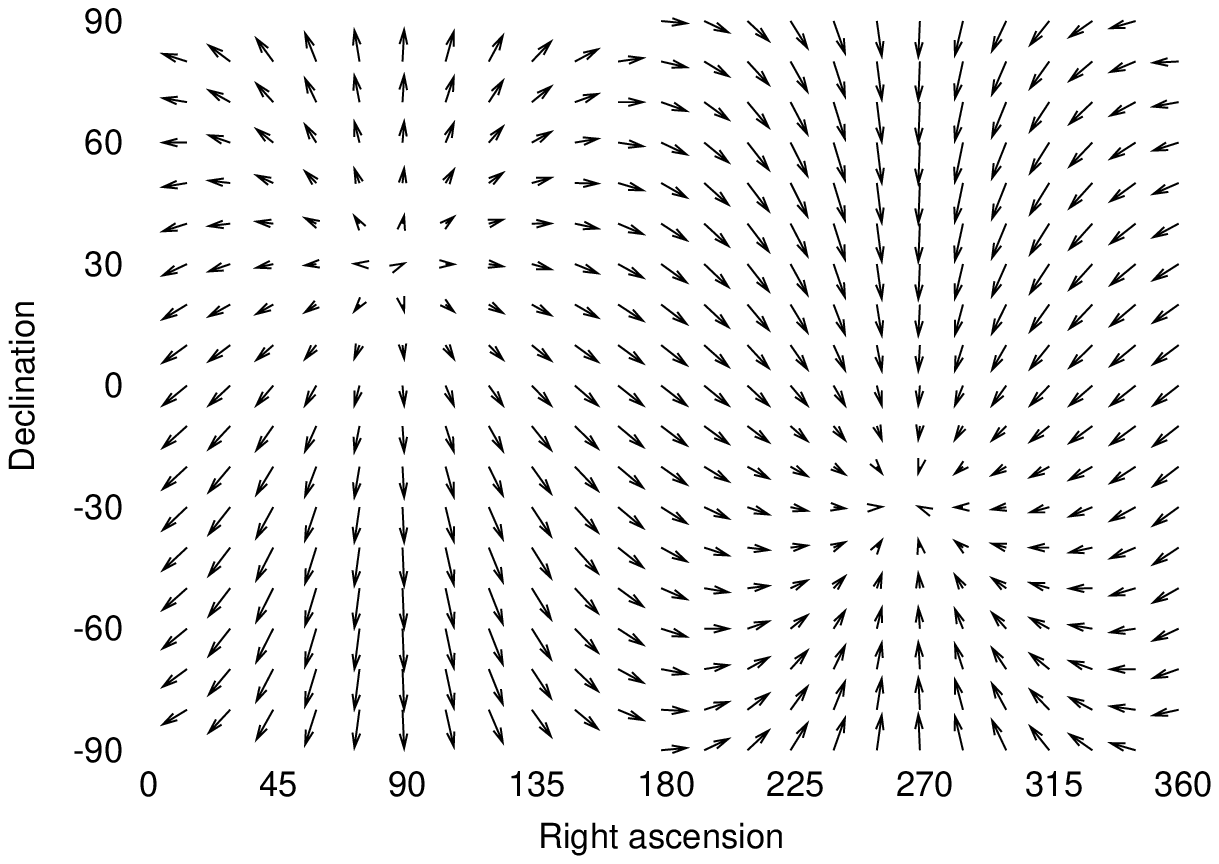}
\caption{Proper motions of sources due to GA in Galactic (left) and equatorial (right) coordinates. Arrows of the maximum 
length correspond to a proper motion of 5 $\mu$as/yr.}
\label{fig:pm_aberr}
\end{figure}

The accuracy with which corrections for GA can 
be calculated depends on the accuracy of A; i.e., the 
accuracy of $R_0$ and $\Omega_0$ or equivalent parameters, such 
as the Oort constants. As the data of the survey 
[13] and the later data [14, 15] show, this accuracy 
is no better than 5\%.
There is also ambiguity in 
the transformation between Galactic and equatorial 
coordinates [16]. All these circumstances means that 
the accuracy of (3), (4) is 5--10\%. 
The continuous accumulation of VLBI observations
 and enhancement of their accuracy leads us to 
pose the question of whether it is possible to refine the 
parameters of the Galactic rotation based on VLBI 
observations of extragalactic radio sources [17, 18]. 
Unfortunately, results obtained by different authors 
applying different methods to different data are somewhat
 contradictory, and yield values of the GA constant
 that can differ appreciably [18--23]. However, 
there is hope that, as new observational material is 
accumulated and VLBI technology is developed, for 
example, as a result of the realization of the next-
generation VLBI network VLBI2010 [24], the accuracy
 of these estimates will grow, and new VLBI 
data together with the results of space astronometric 
measurements will enable the refinement of stellar-
astronomy data over the next ten to fifteen years. 
However, already now, the accuracy of the GA constant
is sufficient to calculate GA with uncertainties 
below 1$\mu$as/yr, which enables reliable estimation of 
possible systematic effects in VLBI observations due 
to GA. Here, we will consider the influence of GA on 
derived precession parameters.

\section{Influence of GA on estimates of precession parameters}

Corrections to the precession parameters based on 
VLBI observations can be derived from analyses of 
series of coordinates of the CP obtained from individual
 24-hour observing sessions carried out on global 
VLBI networks. On average, about three such sessions
 are conducted each week, primarily using networks
 with good geometrical characteristics, making 
it possible to obtain high-accuracy estimates of the 
Earth-rotation parameters, including the coordinates 
of the CP [25]. A brief description of the main observational
 programs engaged in this work is given 
in [26, 27]. 

In practice, VLBI observations measure the offset 
in the position of the CP (Celestial Pole Offset, CPO)

Linear trends and 18.6-year harmonics in the CPO series 
$dX$ and $dY$ representing the difference between
the measured and theoretical coordinates of 
the CP. After correcting for thefreecorenutation 
(FCN), which does not appear in the precession-nutation theory and is modeled as an empirical effect
[28], the CPO measurements are interpreted as 
reflecting errors in the adopted precession-nutation 
model. The observed trend in the series of $dX$ and $dY$ 
measurements can be used to refine the precession 
parameters [10, 11]. However, this trend could also 
be due to errors in taking into account proper motions 
of the observed objects, including those due to GA, 
which depend on the position of the object on the 
celestial sphere. Note that, in the ideal case, when 
using a single set of objects uniformly distributed 
over the celestial sphere over an extended observing 
period, the errors in the precession parameters should 
be zero. However, neither of these conditions are fulfilled
in practice. The distribution of sources in right 
ascension is fairly uniform, apart from some small 
gaps near the Galactic equator, but the same is 
not true of the declination distribution of the sources. 
Because the majority of VLBI stations are located 
in the Northern hemisphere, most of the observed 
radio sources are located in the Northern sky (see, for 
example, [29, 30]). Below, we present additional data 
on the declination distribution of the observed radio 
sources. 

\begin{table}
\centering
\caption{Linear trends and 18.6-year harmonics in the CPO series}
\tabcolsep=4pt
\begin{tabular}{ccc}
\hline
& \begin{tabular}{ccc}
Series & Without & With \\
Series & correction for GA & correction for GA \\
\multicolumn{3}{c}{\phantom{Amplitude of the 18.6-year harmonics, $\mu$as}} \\[-1em]
\end{tabular}
\\
\hline
Version 1
& \begin{tabular}{ccc}
\multicolumn{3}{c}{\phantom{Amplitude of the 18.6-year harmonics, $\mu$as}} \\[-1em]
\multicolumn{3}{c}{Linear trend, $\mu$as/yr} \\
$dX$~~ & $\phantom{-1} 9.6 \pm 0.5$ & $\phantom{-1} 9.7 \pm 0.5$ \\
$dY$~~ &            $-18.8 \pm 0.6$            & $-18.6 \pm 0.6$ \\
\end{tabular}
\\
&&\\
\hline
Version 2
& \begin{tabular}{ccc}
\multicolumn{3}{c}{Linear trend, $\mu$as/yr} \\
$dX$~~ & $\phantom{-1} 3.6 \pm 0.8$ & $\phantom{-1} 3.6 \pm 0.8$ \\
$dY$~~ &            $-14.0 \pm 0.8$            & $-14.0 \pm 0.8$ \\
\multicolumn{3}{c}{Amplitude of the 18.6-year harmonics, $\mu$as} \\
$dX$~~ & $\phantom{-} 60.9 \pm 5.7$ & $\phantom{-} 62.6 \pm 5.7$ \\
$dY$~~ & $\phantom{-} 55.5 \pm 5.5$ & $\phantom{-} 54.7 \pm 5.5$ \\
\end{tabular}
\\
\hline
\end{tabular}
\label{tab:trend}
\end{table}

Here, we consider VLBI observations in the 
database of the International VLBI Service for 
Geodesy and Astrometry (IVS) [26]. Observations 
from 3136 sessions conducted between January 5, 
1984 and March 12, 2010 were reduced, comprising 
5.6 million observations (radio-interferometric delays)
 in all. The data reduction was carried out in 
two ways. The first applied the traditional method, in 
which the coordinates of the radio sources were taken 
to be equal to their values in the ICRF2 catalog [30] 
at all epochs. In the second, the source coordinates 
at the observing epoch were calculated taking into 
account their aberration proper motions using (4). 

\begin{figure}
\centering
\includegraphics[width=0.5\textwidth,clip]{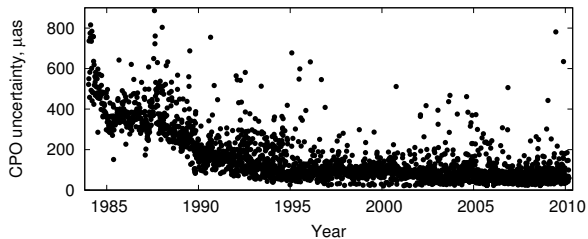}
\caption{Uncertainties in the positions of the CP derived from individual sessions of VLBI observations.}
\label{fig:cpo_err}
\end{figure}

\begin{figure}
\centering
\includegraphics[width=0.5\textwidth,clip]{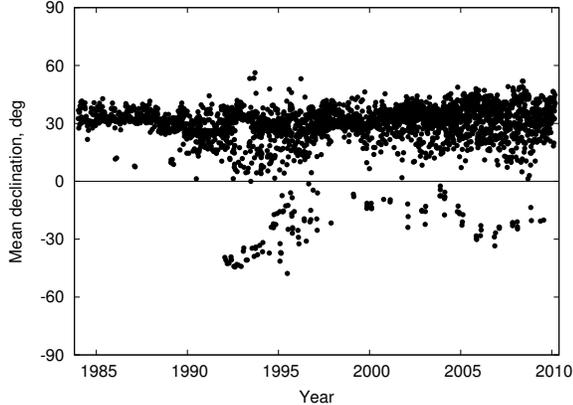}
\caption{Mean declination of radio sources for individual sessions processed in the current study.}
\label{fig:de_mean}
\end{figure}

This yielded two CPO series consisting of 3136 
estimates of $dX$ and $dY$ with median uncertainties 
of 66~$\mu$as. Nineteen measurements were excluded 
from the reduction because they corresponded to 
anomalously large CPO values or had anomalously 
large uncertainties. The noise in both CPO series 
was estimated using a modified version of WMADEV, 
which yields a two-dimensional weighted estimate 
of the noise in a time series [32]. This noise proved 
to be 159~$\mu$as for both series. This value is a mean 
characteristic of the accuracy of CPO values derived 
from VLBI observations obtained over 26 years. As 
was shown in [33], the accuracy of these observations 
grew appreciably over this time period. Figure~2 
presents the revised data for the two-dimensional 
uncertainties in the coordinates of the CP calculated 
as the square root of the squared uncertainties in 
$dX$ and $dY$ . The uncertainties in the CPO were 
reduced appreciably in the first roughly 10 years of 
the observing period, after which they have remained 
essentially the same. 
In addition, the mean declination of the observed 
radio sources for each session is presented in Fig.~3, 
which shows the substantial asymmetry in the declination distribution of these sources. 

We removed the component corresponding to the 
FCN from the 26-year series of $dX$ and $dY$ measurements
 using the ZM2 model [28]. Further, we 
obtained a least-squares fit to calculate the parameters
 of the linear trend, applying weights calculated 
as the inverse squares of the uncertainties in $dX$ and 
$dY$. The calculation of the parameters of the linear 
trend was then repeated simultaneously fitting for the 
parameters of the main nutation term with a period of 
18.6 years. The results are presented in the table. 
Our calculations show that the influence of GA on 
the precession parameters derived from VLBI observations
 could comprise up to 20~$\mu$as per century, depending
 on the reduction methods used. We have also 
determined the influence of GA on the parameters of 
low-frequency nutation terms. 

\section{Conclusion}

We have estimated the influence of GA on the 
parameters of precession models derived from VLBI 
observations, which are currently the main method 
for refining precession-nutation theory. We analyzed 
a series of VLBI measurements covering 26 years, 
consisting of 3126 CPO estimates obtained from 24-hour
global VLBI observing sessions. We compared
the parameters of linear trends in the measured
CP coordinates obtained with and without including
the effect of GA, finding differences reaching 
0.2~$\mu$as/year, or 20~$\mu$as per century. Allowing for 
GA also influences the corrections to the amplitudes 
of long-period nutation terms at the level of one to 
two~$\mu$as, which, in turn, leads to a dependence of 
the coefficients of the linear trend, and thereby of 
the precession parameters, on the refined parameters. 
Note that modern requirements for the accuracy of 
precession theory are 1~$\mu$as per century [10]. 
Thus, our results indicate that the influence of GA 
on estimates of precession parameters based on VLBI 
observations is small but not negligible. Therefore, 
in spite of the fact that this effect is smaller than 
the uncertainty with which it can be determined, it is 
necessary to include the effect of Galactic aberration 
in standard algorithms for the reduction of modern 
astrometric observations. Although current calculated
values of the GA constant are only accurate 
to about 10\%, this is sufficient to take into account 
the main influence of GA. Naturally, increasing this 
accuracy to 1--2\% is very desirable to enable fuller 
correction for this effect in the future. 

\section*{References}

\noindent
\leftskip=\parindent
\parindent=-\leftskip

1. 
A. M. Clerke, The System of the Stars (Longmans, 
Green and Co., London, 1890). 

2. 
S. A. Klioner and M. Soffel, Astron. Astrophys. 334, 
1123 (1998). 

3. 
J. Kovalevsky, Astron. Astrophys. 404, 743 (2003). 

4. 
S. Kopeikin and V. Makarov, Astron. J. 131, 1471 
(2006). 

5. 
J. L. Fanselow, JPL Publ. No. 83-39 (JPL, Pasadena, 
1983). 

6. 
T. M. Eubanks, D. N. Matsakis, F. J. Josties, et 
al., in Astronomical and Astrophysical Objectives 
of Sub-Milliarcsecond Optical Astrometry, Ed. by 
E.~H{\o}g and P.~K.~Seidelmann (Kluwer, Dordrecht, 
1995), p. 283. 

7. 
O. J. Sovers, J. L. Fanselow, and C. S. Jacobs, Rev. 
Mod. Phys. 70, 1393 (1998). 

8. 
GAIA: Composition, Formation and Evolution of 
the Galaxy. Concept and Technology Study Report,
 ESA-SCI 4 (European Space Agency, Noordwijk,
 2000). 

9. 
V. E. Zharov, I. A. Gerasimov, and K. V. Kuimov, in 
Proceedings of 25th IAU General Assembly, Joint 
Discussion 16, Ed. by R.~Gaume, D.~McCarthy, and 
J.~Souchay (USNO, Washington, 2003), p. 141. 

10. 
N. Capitaine, P. T. Wallace, and J. Chapront, Astron. 
Astrophys. 412, 567 (2003). 

11. 
N. Capitaine, P. T. Wallace, and J. Chapront, Astron. 
Astrophys. 432, 355 (2005). 

12. 
O. Titov, Mon. Not. R. Astron. Soc. 407, L46 (2010). 

13. 
T. Foster and B. Cooper, arXiv:1009.3220 (2010). 

14. 
V. V. Bobylev, Astron. Lett. 36, 634 (2010). 

15. 
V. V. Bobylev and A. T. Bajkova, Mon. Not. R. Astron. 
Soc. 408, 1788 (2010). 

16. 
J.-C.Liu, Z. Zhu, and H. Zhang, Astron.Astrophys. 
526, A16 (2011). 

17. 
V. E. Zharov, Spherical Astronomy (Vek-2, 
Fryazino, 2006) [in Russian]. 

18. 
O. Titov, in Proceedings of the Conference on 
Journ\'ees 2007 Syst\`emes de R\'ef\'erence
Spatio-temporels, Ed. by N.~Capitaine (Paris Observatory, 
Paris, 2008), p. 16. 

19. 
O. Titov, in Proceedings of the 5th IVS General 
Meeting, Ed. by A.~Finkelstein and D.~Behrend (IAA 
RAS, St. Petersburg, 2008), p. 265. 

20. 
Z. Malkin and E. Popova, in Proceedings of the Conference
on Journ\'ees 2008 Syst\`emes de R\'ef\'erence 
Spatio-temporels, Ed. by M.~Soffel and N.~Capitaine 
(Paris Observatory, Paris, 2009), p. 239. 

21. 
E. A. Popova, in Proceedings of the All-Russia 
Astrometric Conference Pulkovo-2009, Izv. GAO 
Pulkovo 219 (4), 273 (2009) [in Russian]. 

22. 
O. Titov, in Proceedings of the 19th European VLBI 
for Geodesy and Astrometry Working Meeting, Ed. 
by G.~Bourda, P.~Charlot, and A.~Collioud (Univ. 
Bordeaux 1 -CNRS, Bordeaux, 2009), p. 14. 

23. 
O. Titov, S. B. Lambert, and A.-M. Gontier, Astron. 
Astrophys. 529, A91 (2011). 

24. 
D. Behrend, J. Boehm, P. Charlot, et al., in Observing 
our Changing Earth, Ed. by M.~G.~Sideris, IAG 
Symp. 133 (Springer, 2008), p. 833. 

25. 
Z. Malkin, J. Geod. 83, 547 (2009). 

26. 
W. Schl\"uter and D. Behrend, J. Geod. 81, 379 (2007). 

27. 
Z. M. Malkin, Astron. Rep. 54, 1053 (2010). 

28. 
Z. M. Malkin, Solar Syst. Res. 41, 492 (2007). 

29. 
O. Titov and Z. Malkin, Astron. Astrophys. 506, 1477 
(2009). 

30. 
C. Ma, E. F. Arias, G. Bianko, et al., in The Second 
Realization of the International Celestial Reference
Frame by Very Long Baseline Interferometry,
IERS Techn. Note No.~35, Ed. by A.L.~Fey, 
D.~Gordon, and C.S.~Jacobs (Verlag des Bundesamts 
fuer Kartographie und Geodaesie, Frankfurt am Main, 
2009). 

31. 
Z. Malkin, J. Geod. 82, 325 (2008). 

32. 
Z. M. Malkin, Kinem. Phys. Celest. Bodies, 27, 42 (2011). 

33. 
Z. M. Malkin, Izv. GAO Pulkovo 218, 397 (2006) [in Russian]. 

\vskip 2em
Translated by D. Gabuzda 

\end{document}